\documentclass{svproc}
\usepackage{graphicx}%
\usepackage{multirow}%
\usepackage{amsmath,amssymb,amsfonts}%
\usepackage{mathrsfs}%
\usepackage[title]{appendix}%
\usepackage{xcolor}%
\usepackage{textcomp}%
\usepackage{manyfoot}%
\usepackage{booktabs}%
\usepackage{algorithm}%
\usepackage{algorithmicx}%
\usepackage{algpseudocode}%
\usepackage{listings}%
\usepackage{url}

\def\orcidID#1{\unskip$^{[#1]}$} 

\begin{document}
\mainmatter              
\title{Tracing Money Flows in the Multi-Token Cryptocurrency Network}
%
%
\author{Natkamon Tovanich\inst{1}\orcidID{0000-0001-9680-9282} \and Célestin Coquidé\inst{2}\orcidID{0000-0001-8546-6587}}
\authorrunning{Natkamon Tovanich and Célestin Coquidé} 

\institute{TU Wien, Vienna, Austria\\\email{natkamon.tovanich@tuwien.ac.at} \and UTINAM Institute, UMR 6213 CNRS, Université Marie et Louis Pasteur, Besançon, France \email{coquide31@gmail.com}}

\maketitle              

\begin{abstract}
Tracing money flows in cryptocurrency transaction networks is challenging because funds can move across accounts, transactions, and token layers.
We propose a transaction-aware graph in which nodes represent account--asset states within transactions, while edges encode observed transfers, inferred asset transformations, and temporal inter-transaction sequences.
Applied to Aave borrow--repay flows, we compare graph construction and edge-pruning strategies and extract shortest paths using a heuristic weighting.
Our results reveal a trade-off between reachability and interpretability and uncover frequent token sequences underlying how borrowed assets are transformed before repayment.
This work lays a foundation for general, interpretable multi-token flow tracing.
\end{abstract}
\section{Introduction}
\label{sec:introduction}

Tracing money flows in cryptocurrency transaction networks is increasingly challenging, as smart contract blockchains such as Ethereum enable developers to create new tokens and decentralized finance (DeFi) protocols, creating on-chain ecosystems for trading, lending, and liquidity provision. As these activities grow more complex, strategies involving flash loans, maximal extractable value (MEV), and arbitrage combine multiple protocols and assets, making it difficult to determine how value moves across accounts, transactions, and token layers.


Understanding these flows is important for analyzing how DeFi protocols are used and how participants execute on-chain financial strategies. However, conventional transaction graphs represent accounts as nodes and transfers as edges, making them less suitable for tracking value across multiple tokens and related transaction sequences. To address this gap, we introduce a transaction-aware network representation and compare path finding strategies for tracing multi-token money flows. Our contributions are threefold:

\begin{enumerate}
    \item We model each node as an account--asset state within a transaction and connect these states using observed transfers, inferred asset transformations (e.g., swaps), and temporal inter-transaction relationships.
    \item We compare graph construction and edge pruning strategies and introduce a heuristic weighting scheme that distinguishes directly observed, inferred, and lower-confidence connections.
    \item Using Aave borrow--repay flows as a case study, we characterize how these modeling choices affect path reachability, length, composition, and similarity, and analyze frequent token transformation patterns.
\end{enumerate}

Our results reveal trade-offs among multi-token flow-tracing strategies and illustrate how borrowed assets move across transactions and token layers. These findings lay the groundwork toward a general, interpretable framework for tracing cryptocurrency flows and analyzing on-chain financial activity.




\section{Related Work}
\label{sec:related_work}


Although cryptocurrency flow tracing is central to blockchain forensics, multi-token representations and tracing strategies remain underexplored. Early work has focused on Bitcoin, applying taint-allocation rules such as Poison, Haircut, FIFO, and LIFO~\cite{tironsakkul2019taint}, or representing taint flows as dynamic graphs~\cite{tovanich2023fingerprinting}. For account-based blockchains, TRacer formulates tracing as a graph search over directed, weighted, temporal, and multi-relational networks~\cite{wu2023tracer}.


Cryptocurrency flow is widely used to detect fraudulent and illicit activity. On Bitcoin, MFScope traces funds linked to Dark Web addresses~\cite{lee2019cybercriminal}, while graph learning methods classify illicit Bitcoin transactions~\cite{weber2019aml}. On Ethereum, DenseFlow identifies Ethereum laundering groups and pathways~\cite{lin2024denseflow}, and Wu et al.\ trace heist assets through token swaps and service providers~\cite{wu2024assetflows}. Recent works incorporate transaction semantics for EVM and cross-chain tracing~\cite{fu2026flowshield,huo2025mftracer,wu2026amlguard}. These methods rely on black-box machine learning models to flag suspicious activity from known illicit accounts, whereas our interpretable approach reconstructs multi-token paths, with forensic use as a possible extension.

Ethereum token transfers have also been studied as account-level networks~\cite{coquide2025analysis,loporchio2024comparing,victor2019erc20}. Within DeFi, prior network work has examined protocol compositions~\cite{kitzler2023defi} and used token transfer motifs to identify transaction methods~\cite{tovanich2024decoding}. Studies of decentralized lending investigate motivations behind using Aave~\cite{cornelli2025defilending}, liquidation risks~\cite{qin2021liquidations}, and time to repayments~\cite{green2024defi}.
We complement this literature by connecting transaction-specific account--asset states to trace how borrowed assets are used and transformed between Aave borrowing and repayment events.

\section{Material and Methods}
\label{sec:methodology}

\subsection{Data Acquisition and Preprocessing}

For each ego account and tracing instance, we retrieve the relevant history of native-currency, ERC-20 (fungible), ERC-721 (non-fungible, NFTs), and ERC-1155 (multi-token) transfers from the Alchemy Transfer API~\cite{alchemy_transfers_api}. Each transfer event records the block number, transaction hash, sender, recipient, token contract, asset symbol, transfer category, and the amount transferred.

Before constructing the network, we remove self-loops because they do not represent movement between distinct accounts. Native ETH transfers are relabeled as WETH to represent native and wrapped Ether in a common asset layer, while technical wrapping and unwrapping transfers involving the WETH contract are removed. Missing ERC-721 and ERC-1155 quantities are set to one. For mint and burn events, the zero-address endpoint is replaced with the corresponding token contract address, attributing token creation or destruction to the issuing contract rather than to a shared zero-address node.

Finally, transfers with the same block number, transaction hash, sender, recipient, asset, and token-contract address are aggregated by summing their amounts. This produces a single weighted transfer for repeated movements of the same asset between the same accounts within a transaction.

\subsection{Transaction-Aware Multi-Token Network}


We represent the multi-token flow network as a directed graph $G=(V,E)$, where each node $v=(h,a,\tau)$ denotes a transaction hash $h$, account address $a$, and asset $\tau$. Unlike an account-level graph, this separates the same account across transactions and assets, preserving temporal and asset-specific information. The network combines \emph{within-transaction} and \emph{inter-transaction} edges.

\subsubsection{Within-Transaction Edges} connect token transfers and asset exchanges that occur within the same transaction. We distinguish three edge types:



\paragraph{Transfer edges.} An aggregated token transfer creates a directed edge $\bigl((h,a_s,\tau),\allowbreak(h,a_r,\tau)\bigr) \in E^{\mathrm{tr}}$, where $a_s$ and $a_r$ are the sender and recipient, respectively. These edges represent directly observed transfers, including mint and burn operations, after replacing the zero address with the token contract.

%
%

\paragraph{Swap edges.} A swap edge connects different asset layers when a counterparty receives one asset and sends another within the same transaction. Specifically, when account $a$ receives asset $\tau_{\mathrm{in}}$ and sends a different asset $\tau_{\mathrm{out}}$ in transaction $h$, we add $\bigl((h,a,\tau_{\mathrm{in}}), (h,a,\tau_{\mathrm{out}})\bigr) \in E^{\mathrm{sw}}$, $\tau_{\mathrm{in}}\neq\tau_{\mathrm{out}}$ and $a\neq a_{\mathrm{ego}}$. 
The edge direction represents the assumed conversion from the received asset to the asset subsequently spent. Such edges capture interactions including token swaps, deposits, redemptions, and tokenized claims. They are created only for counterparty accounts, rather than the ego account, to avoid assuming every combination of assets received and spent by the ego represents a direct conversion.


\paragraph{Teleportation edges.} A teleportation edge is used as a fallback when token operations appear related but cannot be connected through an explicit transfer or identifiable swap. Let $G_h^{\mathrm{intra}}$ denote the current intra-transaction graph for transaction $h$, and let $u\not\leadsto_{G_h^{\mathrm{intra}}}v$ indicate that no directed path exists from $u$ to $v$.
For two distinct asset nodes belonging to the same account and transaction, we add $\bigl((h,a,\tau_i),(h,a,\tau_j)\bigr) \in E^{\mathrm{tp}}$, $\tau_i\neq\tau_j$, $\bigl((h,a,\tau_i),(h,a,\tau_j)\bigr) \notin E^{\mathrm{sw}}$, and $(h,a,\tau_i) \not\leadsto_{G_h^{\mathrm{intra}}}(h,a,\tau_j)$.
That is, an edge is added only when no swap edge or intra-transaction path already connects them. 
These edges preserve possible connectivity between related token operations but represent lower-confidence relationships rather than observed transfers or identified economic conversions. They therefore increase reachability without necessarily improving the accuracy of the reconstructed flow.


%
%


\subsubsection{Inter-Transaction Edges} connect receipts and expenditures of the same account and asset across different transactions. For each node $(h,a,\tau)$, we calculate the net balance change
\begin{equation}
\Delta_{h,a,\tau}
=
\sum_{e\in E^{\mathrm{in}}_{h,a,\tau}}x_e
-
\sum_{e\in E^{\mathrm{out}}_{h,a,\tau}}x_e,
\end{equation}
where $E^{\mathrm{in}}_{h,a,\tau}$ and $E^{\mathrm{out}}_{h,a,\tau}$ are the incoming and outgoing transfer edges of the node and $x_e$ is the transferred amount. Mint edges are excluded from the outgoing amount, and burn edges from the incoming amount. Consequently, $\Delta_{h,a,\tau}>0$ indicates a receipt, $\Delta_{h,a,\tau}<0$ an expenditure, and nodes with $\Delta_{h,a,\tau}=0$ receive no inter-transaction edges.

%
%
%
%


Connecting each receipt to every subsequent expenditure would create a dense set of temporally valid but increasingly implausible edges. 
Inspired by the FIFO and LIFO allocation conventions used in taint tracking~\cite{tironsakkul2019taint}, we apply a recency-based cumulative-expenditure cutoff, referred to as the \emph{LIFO-cutoff}, which approximates rather than implements LIFO accounting.

%
%

For each receipt $r_i=\Delta_{h_i,a,\tau}>0$, let $h_{j_1},\ldots,h_{j_{m}}$ denote the subsequent transactions with expenditures $\Delta_{h_{j_\ell},a,\tau}<0$, ordered chronologically. We connect the receipt node $(h_i,a,\tau)$ to the expenditure nodes $(h_{j_\ell},a,\tau)$ for $\ell\leq c_i^{*}$, where
%
%
\begin{equation}
c_i^{*} = \min\left\{k\middle|\, \sum_{\ell=1}^{k}\left|\Delta_{h_{j_\ell},a,\tau}\right| \geq r_i\right\}.
\end{equation}
If the cumulative expenditure never reaches $r_i$, all $m$ expenditures are retained ($c_i^{*}=m$). Thus, each receipt is connected only to the earliest subsequent expenditures whose cumulative value is sufficient to exhaust the received amount.

This LIFO-cutoff rule assumes received funds are spent soon after receipt, rather than after older holdings. It favors short temporal spans and avoids reconstructing each account's complete balance history, making it suitable for ego-based tracing that begins from a specific borrowing or transfer event, though it remains an accounting approximation rather than a definitive attribution of fungible tokens.
The all edges variants drop the recency assumption, keeping only temporal order.








\subsection{Aave Borrow-to-Repay Matching}

We demonstrate our methodology with Aave V3, a leading decentralized lending protocol. Aave allows users to supply assets to liquidity reserves and borrow against collateral. It is a suitable use case because borrowed funds may pass through other accounts, assets, and protocols before repayment, and its \texttt{Borrow} and \texttt{Repay} events provide identifiable anchors for defining tracing instances.

For each borrowing event $\beta_i$, we identify chronologically subsequent repayments by the same borrower and reserve asset. The borrower is identified via the \texttt{onBehalfOf} field of \texttt{Borrow} and the \texttt{user} field of \texttt{Repay}. 
%
%

Let $q_i^{\mathrm{B}}$ denote the borrowed amount and $\rho_{j_1},\ldots,\rho_{j_{{m}}}$ the ordered subsequent repayments with amounts $q_{j_\ell}^{\mathrm{R}}$. We retain repayments $\rho_{j_1},\ldots,\rho_{j_{k_i^*}}$, where
%
%
\begin{equation}
k_i^{*}=\min\left\{k\,\middle|\, \sum_{\ell=1}^{k}q_{j_\ell}^{\mathrm{R}} \geq q_i^{\mathrm{B}}\right\}.
\end{equation}
If the cumulative repayment remains below the borrowed amount, all $m$ repayments are retained ($k_i^{*}=m$). This recency-based matching selects the earliest sufficient set of subsequent repayments, yielding the shortest observed temporal span from borrowing to repayment. 

For each borrowing event with at least one retained repayment, we construct the transaction graph from the borrowing block to the block of the final retained repayment. The origin is the transaction--user--asset node representing receipt of the borrowed asset: $s_i=(h_i^{\mathrm{B}},a_i^{\mathrm{user}},\tau_i)$, and the target set contains the \emph{Aave Pool} node for every retained repayment:
$\mathcal{D}_i = \left\{(h_{j_\ell}^{\mathrm{R}},a^{\mathrm{pool}},\tau_i) \,\middle|\, 1\leq\ell\leq k_i^{*} \right\}$.



\subsection{Shortest Path Tracing}
\label{sec:weighted-shortest-path}

For each borrow--repay pair, we extract a single path from the borrowing node $s$ to each repayment node $d$ under two settings: (i) \emph{unweighted} $P_{s,d}^{\mathrm{hop}}$, which minimizes the number of edges, and (ii) \emph{weighted} $P_{s,d}^{\mathrm{w}}$, which minimizes the total traversal cost. The weighting scheme is defined as follows.

\subsubsection{Within-transaction weighting.} Within a transaction, we favor paths that follow asset transfers or exchanges, while using teleportation only as a fallback when no direct path exists.
\begin{itemize}
    \item Transfer edges are assigned a low cost of 0.1, as directly observed token movements.
    \item Swap edges receive an even lower cost of 0.01 to encourage inferred conversions between received and spent assets.
    \item Teleportation edges are assigned a substantially higher cost of 100 as they represent connections without an explicit transfer or an identifiable swap.
\end{itemize}


Furthermore, an inter-transaction edge directly connecting the borrowing node to the ego-account state in a repayment transaction is assigned a cost of 100 to prevent the path from bypassing the intermediate fund movements under investigation.

\subsubsection{Inter-transaction weighting.} We prefer temporally closer expenditures within the LIFO-cutoff and discourage shortcuts that bypass intermediate transactions. An edge within the cutoff costs 1, whereas an edge beyond it costs 10. This penalty steers the path toward the earliest plausible subsequent expenditure rather than a distant one.

Edges involving alter accounts incur an additional cost of 50 (51 within the cutoff, 60 beyond), prioritizing ego-account continuity, since alter accounts' histories are only partially observed and could otherwise create artificial shortcuts (e.g., a deposit followed by a later withdrawal).

\bigskip
\noindent These weights are heuristic traversal costs rather than probabilities or empirically calibrated confidence scores. They were set manually to encode an ordinal preference with roughly order-of-magnitude gaps, so rank order matters more than exact values. The weighted path should be read as the one that best satisfies our preference hierarchy, rather than as definitive evidence that the traced funds followed that exact route. 




\section{Results}
\label{sec:results}





We analyze 1,505 borrower accounts on Aave v3 on the Ethereum mainnet, each with a cumulative borrowed amount of at least 1 million USD between the protocol’s launch on January 27, 2023, and April 9, 2026. The dataset comprises 175,601 borrow events and 288,764 borrow--repay matches identified using recency-based matching, which we refer to as the \emph{LIFO-cutoff}.

Four inter-transaction network variants are evaluated: (1)~all accounts and all candidate edges, (2)~only ego-account edges, (3)~all accounts with LIFO-cutoff pruning, and (4) ego-account edges with LIFO-cutoff pruning, each without and with teleportation edges. For every configuration, we compare the unweighted shortest path $P_{s,d}^{\mathrm{hop}}$ with the weighted shortest path $P_{s,d}^{\mathrm{w}}$ to examine how edge pruning and weighting affect reachability, path length, and path composition.

\subsection{Path Reachability and Shortest Path Lengths}

\begin{table}[t]
    \centering
    \caption{Percentage of borrow events and borrow--repay pairs connected by a path, path length statistics, and percentage of paths containing at least one teleport edge (\% Tele.) for the shortest path ($P_{s,d}^{\mathrm{hop}}$) and weighted shortest path ($P_{s,d}^{\mathrm{w}}$), across teleport and inter-transaction edge conditions.}
    \label{tab:shortest_path}
    \resizebox{\linewidth}{!}{%
        \begin{tabular}{ll cc cccc cccc}
        \toprule
        & & \multicolumn{2}{c}{\% Path Found} & \multicolumn{4}{c}{$P_{s,d}^{\mathrm{hop}}$} & \multicolumn{4}{c}{$P_{s,d}^{\mathrm{w}}$} \\
        \cmidrule(lr){3-4} \cmidrule(lr){5-8} \cmidrule(lr){9-12}
        Teleport & Inter-Transaction Edges & Borrow & Borrow--Repay & Median & Mean & Std & \% Tele. & Median & Mean & Std & \% Tele. \\
        \midrule
        \multirow{4}{*}{No teleport}
         & All accounts, all edges   & 85.91 & 81.62 & 3.00 & 4.13 & 2.18  & --    & 7.00 & 7.57  & 4.55  & --    \\
         & All accounts, LIFO-cutoff & 57.87 & 47.15 & 5.00 & 7.16 & 12.34 & --    & 9.00 & 10.41 & 13.12 & --    \\
         & Ego account, all edges    & 79.27 & 74.95 & 3.00 & 5.64 & 4.27  & --    & 3.00 & 7.07  & 4.95  & --    \\
         & Ego account, LIFO-cutoff  & 42.42 & 30.70 & 9.00 & 9.68 & 5.37  & --    & 9.00 & 10.05 & 5.25  & --    \\
        \midrule
        \multirow{4}{*}{Teleport}
         & All accounts, all edges   & 99.97 & 98.57 & 3.00 & 3.93 & 1.07  & 30.23 & 7.00 & 7.18  & 4.26  & 17.22 \\
         & All accounts, LIFO-cutoff & 97.29 & 89.66 & 5.00 & 5.40 & 7.13  & 71.29 & 5.00 & 8.01  & 8.59  & 47.98 \\
         & Ego account, all edges    & 99.91 & 97.14 & 3.00 & 3.93 & 1.05  & 47.48 & 5.00 & 6.67  & 4.46  & 22.84 \\
         & Ego account, LIFO-cutoff  & 95.93 & 84.89 & 5.00 & 4.85 & 0.95  & 91.68 & 5.00 & 6.96  & 4.12  & 63.84 \\
        \bottomrule
        \end{tabular}
    }
\end{table}


First, we examine reachability between borrow and repay nodes. Table~\ref{tab:shortest_path} reports the share of borrow events with at least one reachable repayment (``Borrow'') and of individual borrow--repay pairs connected by a path (``Borrow--Repay'') across the eight conditions. 
Teleport edges yield the best reachability, with nearly 100\% of paths found in both metrics, although this reflects more permissive connectivity rather than validated correctness. 
Under the least restrictive edge set (all accounts, all edges), borrow-event reachability rises from 85.91\% to 99.97\% and borrow--repay reachability from 81.62\% to 98.57\% relative to no teleportation.
This comes at the cost of unverified connections: teleport edges appear in 30.23--91.68\% of unweighted but only 17.22--63.84\% of weighted paths.



LIFO-cutoff has the opposite effect, substantially reducing reachability by removing candidate temporal links. Under the all-accounts condition, borrow--repay reachability drops from 81.62\% to 47.15\% without teleportation and from 98.57\% to 89.66\% with it. This drop is expected by construction, since the cutoff graph is a subgraph of the all edges graph. It measures sensitivity to the cutoff rather than its validity, because it cannot distinguish flows that violate the assumed ordering from funds that leave the observed accounts or time window. 
Restricting inter-transaction edges to the ego account further reduces reachability, though more modestly. The combined restriction (no teleport, ego account, LIFO-cutoff) yields the worst reachability, at 42.42\% and 30.70\% for borrow and borrow--repay paths, respectively, compared to 57.87\% and 47.15\% with all accounts. Teleportation can thus bridge otherwise disconnected components, while ego-only restriction and the LIFO-cutoff both reduce reachability.




Table~\ref{tab:shortest_path} also shows a consistent divergence in path length. Weighted shortest paths are longer than unweighted ones in every configuration. For instance, under no teleport, all accounts, all edges, the mean rises from 4.13 ($P_{s,d}^{\mathrm{hop}}$) to 7.57 ($P_{s,d}^{\mathrm{w}}$) as expected, since the weighting scheme favors additional low-cost transfer and swap hops over fewer, costlier inter-transaction or teleport hops. Teleportation, by contrast, shortens and stabilizes unweighted paths (mean 4.13 to 3.93, std.\ 2.18 to 1.07 under all-accounts, all-edges) by providing direct shortcuts in place of long fallback detours, whereas LIFO-cutoff lengthens and destabilizes them (mean 4.13 to 7.16, std.\ 2.18 to 12.34 under all accounts, no teleport) by forcing longer, more circuitous paths when no direct chronological match exists. Because these statistics cover only reachable pairs, cross-configuration comparisons also reflect changes in which pairs remain connected.


These results reveal a trade-off between reachability and path construction assumptions. Teleport edges maximize reachability at the cost of unverified connections, whereas ego-only linkage and the LIFO-cutoff reduce reachability but avoid permissive or partially observed links at the cost of coverage. Weighting balances the two by keeping teleportation as a fallback while favoring lower-cost transfer, swap, and inter-transaction edges whenever available.

\subsection{Shortest Path Distribution}
\begin{figure}[t]
    \centering
    \includegraphics[width=\linewidth]{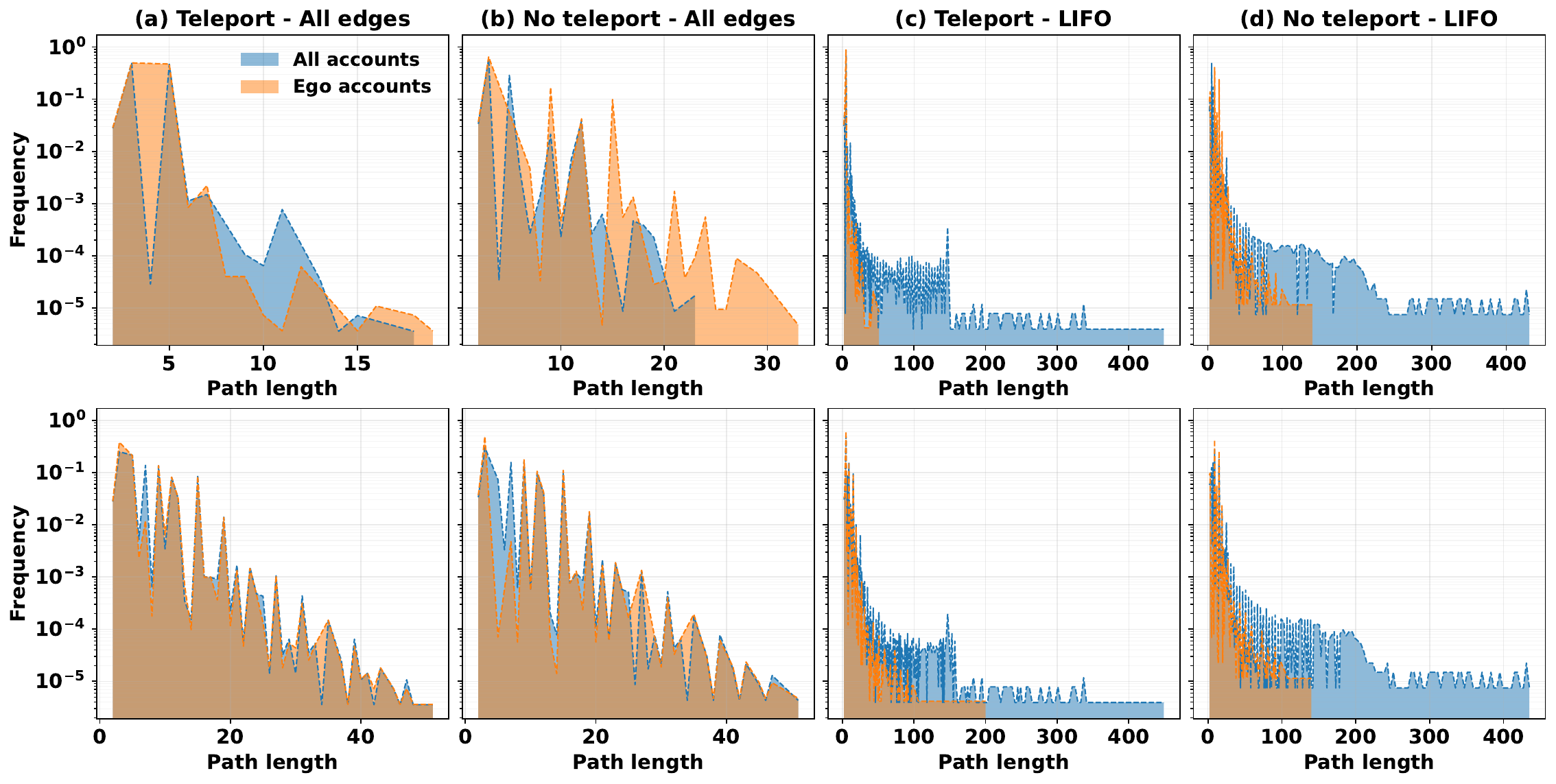}
    \caption{Distribution of shortest path lengths in case of unweighted (top rows) and weighted (bottom rows) scheme for each inter-transaction and teleport conditions.}
    \label{fig:path_length_distr}
\end{figure}

We report the shortest path length distributions obtained under different inter-transaction conditions in Fig.~\ref{fig:path_length_distr}. Experiments conducted under unweighted and weighted schemes are reported in the top and bottom panels, respectively.

Without edge weighting, distributions obtained without LIFO-cutoff are relatively similar across node restrictions, concentrated over a small range: path lengths stay below 20 with teleport, and increase only slightly without it, reaching 23 for all nodes and 33 under ego-node restriction. In contrast, LIFO-cutoff produces markedly different distributions depending on node restriction: with no restriction, path lengths can exceed 400, whereas restricting inter-transaction links to ego accounts sharply reduces the maximum to 140 without teleportation and to 51 with teleportation.



Edge weighting increases similarity between distributions when the LIFO-cutoff is absent, muting the effect of node restriction and extending lengths up to 51 under both teleportation conditions. For LIFO-cutoff paths, weighting has a limited effect without teleportation, but with teleportation and ego-node restriction combined, it substantially extends observed lengths, up to 200.


The impact of node restriction is thus strongly conditioned by the inter-transaction filtering scheme. Edge weighting tends to stabilize path-length distributions across teleportation conditions, suggesting that the resulting paths are less sensitive to the teleportation parameter.

\subsection{Shortest Path Composition}

\begin{figure}[t]
    \centering
    \includegraphics[width=\linewidth]{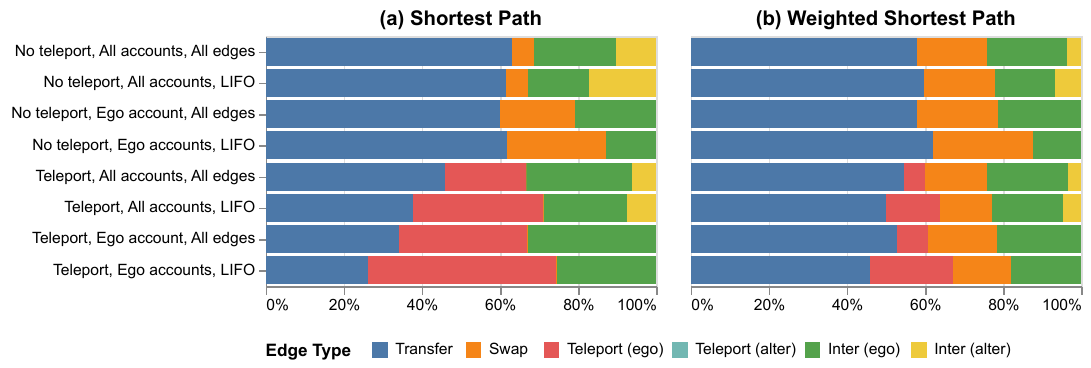}
    \caption{Composition of shortest paths by edge type. Percentages are calculated over all edge occurrences in the shortest paths within each condition.}
    \label{fig:path_composition}
\end{figure}



We next examine edge composition. Fig.~\ref{fig:path_composition}(a) shows transfer edges dominate unweighted, non-teleport paths (59.87\%--63.01\% of edge occurrences). Swap edges account for only 5.62\%--5.67\% when alter-account inter-transaction edges are allowed, which increases to 19.26\% (all edges) and 25.32\% (LIFO-cutoff) when restricted to ego-account edges, suggesting alter-account temporal links offer shorter alternatives that bypass within-transaction walks. Teleport edges reshape this composition further, accounting for 20.75\% of edges (all-accounts, all-edges) up to 47.98\% (ego-only, LIFO-cutoff), while swap edges fall below 0.14\% in every teleport condition. The results show the shortest path uses teleportation as a shortcut rather than longer sequences on swap and transfer edges. This reliance increases as LIFO-cutoff or ego-only restrictions remove alternative inter-transaction connections.



Weighting produces a markedly different composition (Fig.~\ref{fig:path_composition}(b)). Without teleportation, it raises the swap share from 5.62\% to 17.89\% (all accounts, all edges) and from 5.67\% to 18.23\% (LIFO-cutoff). In contrast, alter-account inter-transaction edges fall from 10.30\% to 3.78\% and from 17.34\% to 6.63\%, respectively, reflecting the low cost of swap edges and the added penalty on alter-account edges. Weighting also reduces teleportation reliance: The ego-account teleport share drops from 20.75\%--47.98\% (unweighted) to 5.35\%--21.24\% (weigh-\\ted), while swap edges rise from near zero to 13.37\%--17.68\%. Teleport share remains highest under ego-only and LIFO-cutoff, indicating that no lower-cost alternative exists for some recovered pairs.


Overall, weighting retains teleportation as a fallback while favoring transfer, swap, and ego-account inter-transaction edges, resulting in longer path lengths. Since configurations differ in reachability and thus in which paths are recovered, these compositions should be read relative to each condition's own path set.

\subsection{Shortest Path Similarities}


Furthermore, we examine whether different configurations recover the same paths for a given borrow--repay pair, using Jaccard similarity between node sets (1 = identical nodes, 0 = no overlap). The metric captures path overlap but not the order in which nodes are traversed.

\begin{figure}[t]
    \centering
    \includegraphics[width=\linewidth]{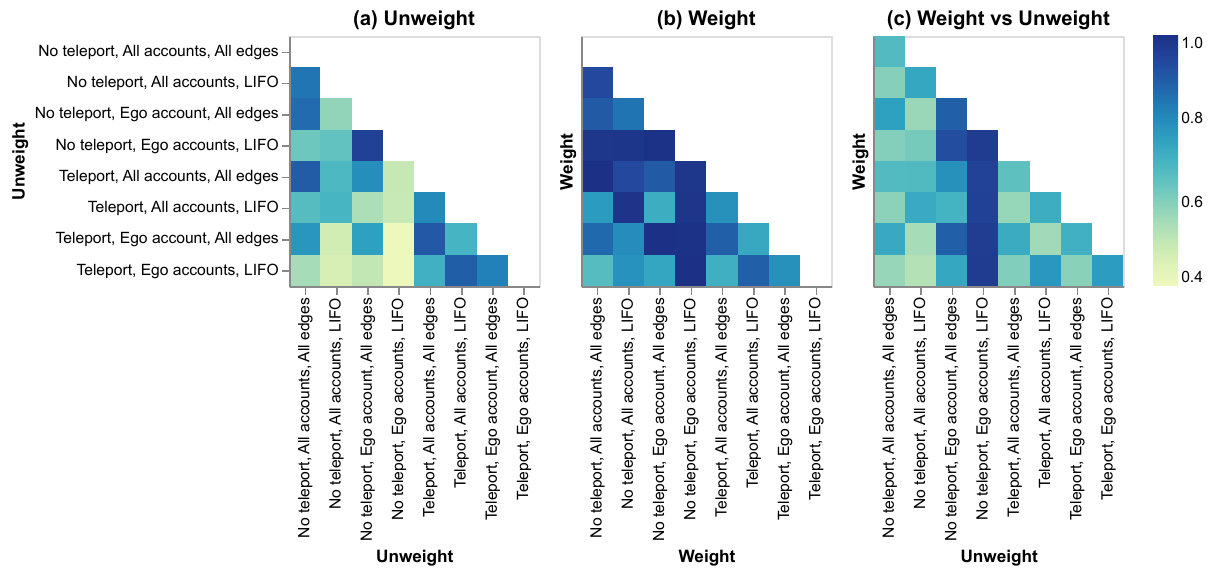}
    \caption{Mean Jaccard similarity between shortest paths obtained under different configurations: (a) unweighted, (b) weighted, and (c) weighted versus unweighted paths.}
    \label{fig:similarity}
\end{figure}


Fig.~\ref{fig:similarity}(a) compares unweighted paths, with mean similarities from 0.408 to 0.957 and the highest between the no-teleport, ego-only configurations with and without LIFO-cutoff (0.957). Adding teleport edges changes unweighted paths more as the inter-transaction network grows more restrictive: similarity with vs.\ without teleportation is 0.896 (all accounts, all edges), 0.755 (ego-only, all edges), 0.694 (all accounts, LIFO-cutoff), and 0.408 (ego-only, LIFO-cutoff).


Weighted paths are markedly more stable (Fig.~\ref{fig:similarity}(b)), with mean similarities from 0.671 to nearly 1.000. Paths with and without teleportation are nearly identical under matching inter-transaction conditions (0.992--1.000), and the no-teleport, ego-only, LIFO-cutoff configuration is highly similar to all four teleport configurations (0.984--1.000), indicating the high teleportation penalty keeps the weighted search from using teleport edges when cheaper alternatives exist.



Comparing weighted against unweighted paths (Fig.~\ref{fig:similarity}(c)), the unweighted no-teleport, ego-only, LIFO-cutoff paths closely match the weighted paths under every teleport configuration (0.954--0.970), suggesting weighting steers the search toward the ego-only, LIFO-cutoff path structure while retaining teleport edges' added reachability. Along the diagonal, where the graph configuration is held constant, weighted vs. unweighted similarity is highest without teleportation under ego-only edges (0.888 all edges, 0.970 LIFO-cutoff) but much lower under teleport configurations (0.658--0.762), confirming that weighting changes paths that would otherwise rely on teleport shortcuts.



\subsection{Frequent Shortest Paths}


Finally, using the teleport-enabled, all-accounts, all-edges configuration, we analyze frequent token sequence patterns in weighted shortest paths to understand how borrowed assets are transformed before repayment. 

\begin{figure}[t]
    \centering
    \includegraphics[width=\linewidth]{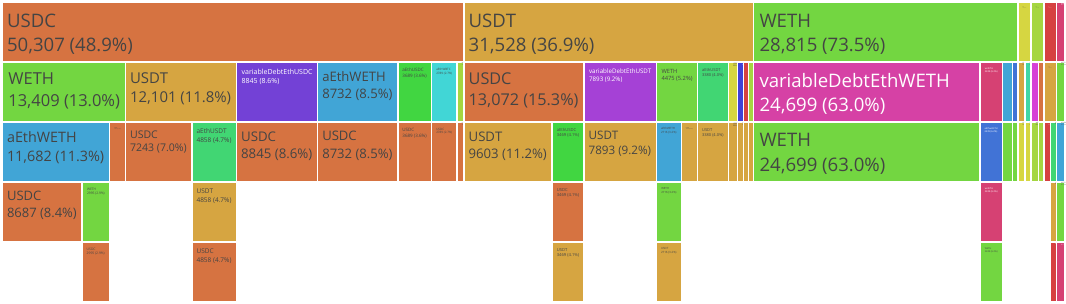}
    \caption{Frequent token-sequence patterns among weighted shortest paths. Rectangle size represents the number of borrow--repay paths. The percentages are calculated relative to the total number of paths. Because direct paths and patterns that occur fewer than 500 times are omitted, the percentages in the figure do not sum to 100\%.}
    \label{fig:frequent_patterns}
\end{figure}

For each shortest path, we strip transaction hashes and account addresses, retaining only the ordered sequence of tokens. Direct repayment without token changes is the most common path (46.88\%), but there is little information on the use of borrowed assets. Thus, we exclude it and focus on paths involving more than one token in the sequence. Fig.~\ref{fig:frequent_patterns} presents an icicle plot of patterns occurring in at least 500 borrow--repay pairs, covering 41.35\% of all paths.


The most frequent interpretable patterns are relatively simple \emph{round-trip swaps} or \emph{Aave supply--withdraw cycles}.
Round-trip swaps, where borrowers swap the borrowed asset for another token and back before repayment, involve $\texttt{USDT} \rightarrow \texttt{USDC} \rightarrow \texttt{USDT}$ (3.4\%) and $\texttt{USDC} \rightarrow \texttt{USDT} \rightarrow \texttt{USDC}$ (2.6\%).
Supply--withdraw cycles, in which depositing the borrowed asset to Aave and withdrawing it for repayment, include $\texttt{USDC} \rightarrow \texttt{aEthUSDC} \rightarrow \texttt{USDC}$ (3.1\%) and $\texttt{USDT} \rightarrow \texttt{aEthUSDT} \rightarrow \texttt{USDT}$ (1.2\%). 
Longer sequences combine both patterns, indicating the borrowed asset is swapped, supplied to Aave for an \texttt{aToken}, withdrawn, and exchanged back before repayment, e.g., $\texttt{USDC} \rightarrow \texttt{USDT} \rightarrow \texttt{aEthUSDT} \rightarrow \texttt{USDT} \rightarrow \texttt{USDC}$ (1.7\%) and the same pattern for borrowed USDT swapped with USDC (1.2\%).



However, the frequent patterns also reveal shortcuts that should not be read as economic strategies. Many shortest paths route through debt-position tokens as an intermediate---most prominently $\texttt{WETH} \rightarrow \texttt{variableDebtEthWETH} \rightarrow \texttt{WETH}$ (8.8\%), and similarly for USDC (3.2\%) and USDT (2.8\%) because no valid shortest path exists otherwise, so the weighted search falls back on this shortcut. 
Other patterns skip expected intermediate steps: $\texttt{USDC} \rightarrow \texttt{WETH} \rightarrow \texttt{aEthWETH} \rightarrow \texttt{USDC}$ (3.1\%) omits an explicit return from aEthWETH to WETH, while $\texttt{USDC} \rightarrow \texttt{aEthWETH} \rightarrow \texttt{USDC}$ (3.1\%) bypasses both WETH conversion steps entirely. Such patterns may arise from fallback teleport edges or unresolved multi-contract interactions; whether they have a systematic cause remains an open question and a priority for future work.


\section{Discussion and Future Work}
\label{sec:discussion}



We introduced a transaction-aware representation for tracing multi-token flows and compared shortest path configurations on Aave borrow--repay flows. The results reveal a trade-off between reachability and interpretability: teleportation edges recover more paths but may introduce implausible shortcuts, whereas restricting inter-transaction edges reduces coverage. Reachability should thus not imply reconstruction accuracy. Heuristic weighting balances these effects by retaining uncertain edges as fallbacks while favoring transfers, swaps, and temporally proximate links. Frequent token patterns reveal round-trip swaps and supply--withdraw cycles, but also artifacts from debt-token shortcuts and missing asset transformations.


Our approach has some limitations. A shortest path cannot capture flows that split, merge, or form cycles, and it may bypass intermediate transactions by linking borrowing directly to repayment.
Since no ground truth exists (only the borrow and repay events are observed, while the route between them is inferred), we assess structural properties and pattern interpretability, treating the paths as hypotheses rather than verified trajectories.
Future work could apply $k$-shortest paths, flow decomposition, and cycle detection to recover more complete strategies, and use sensitivity analysis over the weights and alternative allocation rules (e.g., FIFO, Haircut) to quantify the uncertainty from inferred edges and heuristic weights.
As a partial sanity check, individual conversion steps could be validated against event logs (e.g., supply, withdrawal, and swap events). Beyond event logs, synthetic DeFi transaction data with known ground truth could be generated to evaluate the accuracy of the entire set of inferred paths.
To better characterize borrowers' trading strategies, we plan to enrich the network model by representing each transaction as a hypergraph and incorporating account and protocol labels, transaction semantics, and inter-transaction timing. 
More broadly, we aim to extend the approach beyond Aave by adding cycle-aware tracing and path uncertainty analysis to study evolving DeFi use cases such as on-chain trading, liquidity management, and arbitrage.



\subsubsection*{Acknowledgements}
We thank Adam Hamimed for his preliminary analysis that informed this work, to Julien Prat for motivating the research question and case study, and to Louis Latournerie for providing the Aave v3 dataset.

\bibliographystyle{spmpsci} 
\bibliography{refs} 

\end{document}